\documentclass[twocolumn,showpacs,prc,amsmath,amssymb,superscriptaddress]{revtex4}

\usepackage{graphicx}
\usepackage{dcolumn}
\usepackage{bm}
\usepackage{epsfig}

\begin{document}

\preprint{}

\title{Production of new neutron-rich isotopes of heavy elements in fragmentation reactions of $^{238}$U projectiles at 1 A GeV.}

\author{H.~Alvarez-Pol}
\email{hector.alvarez@usc.es}
\affiliation{Universidade de Santiago de Compostela, 15782 Santiago de 
Compostela, Spain}

\author{J.~Benlliure}%
\affiliation{Universidade de Santiago de Compostela, 15782 Santiago de 
Compostela, Spain} 

\author{L.~Audouin}%
\affiliation{IPN, IN2P3-CNRS, Universit\'e Paris-Sud 11, UMR 8608, F-91406 Orsay, France} 

\author{E.~Casarejos}%
\altaffiliation[Present address: ]{Universidade de Vigo, 36310 Vigo, Spain}
\affiliation{Universidade de Santiago de Compostela, 15782 Santiago de 
Compostela, Spain}

\author{D.~Cortina-Gil}%
\affiliation{Universidade de Santiago de Compostela, 15782 Santiago de 
Compostela, Spain} 

\author{T.~Enqvist}%
\altaffiliation[Present address: ]{CUPP project, P.O. Box 22, 86801 Pyh\"asalmi, Finland}
\affiliation{GSI Helmholtzzentrum f\"ur Schwerionenforschung GmbH, Planckstr. 1,
D-64291 Darmstadt, Germany}

\author{B.~Fernandez}%
\affiliation{Universidade de Santiago de Compostela, 15782 Santiago de Compostela, Spain} 

\author{A.R.~Junghans}%
\affiliation{Forschungszentrum Dresden-Rossendorf, D-01328 Dresden, Germany} 

\author{B.~Jurado}%
\affiliation{Universit\'e Bordeaux I, CNRS/IN2P3, CENBG, BP 120, F-33175 Gradignan, France} 

\author{P.~Napolitani}%
\affiliation{IPN, IN2P3-CNRS, Universit\'e Paris-Sud 11, UMR 8608, F-91406 Orsay, France} 
\affiliation{GSI Helmholtzzentrum f\"ur Schwerionenforschung GmbH, Planckstr. 1,
D-64291 Darmstadt, Germany}

\author{J.~Pereira}
\altaffiliation[Present address: ]{NSCL, Michigan State University, East Lansing, Michigan 48824, USA}
\affiliation{Universidade de Santiago de Compostela, 15782 Santiago de Compostela, Spain}

\author{F.~Rejmund}%
\affiliation{GANIL CEA/DSM-CNRS/IN2P3, BP 55027, F-14076 Caen, France} 

\author{K.-H.~Schmidt}
\affiliation{GSI Helmholtzzentrum f\"ur Schwerionenforschung GmbH, Planckstr. 1,
D-64291 Darmstadt, Germany}

\author{O.~Yordanov}
\altaffiliation[Present address: ]{INRNE, 72 Tzarigradsko chausee, BG-1784 Sofia, Bulgaria}
\affiliation{GSI Helmholtzzentrum f\"ur Schwerionenforschung GmbH, Planckstr. 1,
D-64291 Darmstadt, Germany}

\date{\today}

\begin{abstract}
The production of heavy neutron-rich nuclei has been investigated using cold fragmentation reactions of $^{238}$U projectiles at relativistic energies. The experiment performed at the high-resolving-power magnetic spectrometer FRS at GSI allowed to identify 45 new heavy neutron-rich nuclei: $^{205}$Pt, 
$^{207-210}$Au, $^{211-216}$Hg, $^{213-217}$Tl, $^{215-220}$Pb, $^{219-224}$Bi, $^{221-227}$Po, $^{224-229}$At, $^{229-231}$Rn and $^{233}$Fr. The production cross sections of these nuclei were also determined and used to benchmark reaction codes that predict the production of nuclei far from stability.
\end{abstract}

\maketitle

The possibility to extend the present limits of the chart of the nuclides provides unique opportunities for investigating the nuclear many-body system with extreme values of isospin and most of the stellar nucleosynthesis processes leading to the production of the heaviest elements in our Universe \cite{RNB8}. This is the reason why presently, several new-generation in-flight radioactive-beam facilities are being commissioned, built or designed \cite{RIBF,FAIR,FRIB}. These facilities will take advantage of two reaction mechanisms, fission and fragmentation, for producing nuclei far from stability. Since a large fraction of nuclei at the proton drip-line have already been produced at the present facilities, the new ones will mostly contribute to enlarge our field of action in the neutron-rich side of the nuclear chart.

Fragmentation reactions of $^{48}$Ca beams have been used to produce light neutron-rich nuclei \cite{Moc06} and reach the heaviest known nuclei at the neutron drip-line \cite{Bau07}. Fission reactions have proven to be an optimum reaction mechanism for the production of medium-mass neutron-rich nuclei \cite{Ber97,Arm04}. Recently, intense beams of $^{238}$U at the new ``Radioactive Ion Beam Factory'' in RIKEN made possible the production of 45 new medium-mass neutron-rich nuclei in in-flight fission reactions \cite{Ohn10}. The next step could be the fragmentation of beams of neutron-rich fission fragments such as $^{132}$Sn to produce extremely neutron-rich nuclei as proposed in \cite{Hel03}.

The access to the north-east region of the chart of nuclides seems, however, a real challenge. Indeed, the heaviest known isotopes in this region are still located relatively close to the $\beta$ stability line. A few years ago, it was proposed to use fragmentation reactions of heavy stable projectiles such as $^{238}$U or $^{208}$Pb at relativistic energies to populate that region of the chart of nuclides \cite{Ben99,Ben04}. The idea behind was to take advantage of the large fluctuations in N/Z and excitation energy of the projectile pre-fragments produced in the abrasion stage of the reaction. Those fluctuations should be sufficient to populate cold-fragmentation reaction channels where the incident projectiles lose mostly protons while the excitation energy gained in the process is rather low. The extreme case for these reactions are the proton-removal channels where the projectiles lose only protons, and the excitation energy gained is below the particle-evaporation threshold.

In order to test this idea, we performed an experiment at GSI Darmstadt where a $^{238}$U beam was accelerated at 1 A GeV with the SIS synchrotron with an intensity around 10$^8$ ions/s. The fragmentation residues produced in collisions with a 2500 mg/cm$^2$ beryllium target were analysed with the high-resolving-power magnetic spectrometer Fragment Separator (FRS) \cite{Gei92}. This is a zero-degree magnetic spectrometer with two symmetric sections in order to preserve the achromatism of the system. The spectrometer is characterised by a resolving power B$\rho$/$\Delta($B$\rho$) $\approx$ 1500, a momentum acceptance $\Delta$p/p $\approx$ 3\% and an angular acceptance around its central trajectory $\Delta \theta \approx$ 15 mrad. A profiled energy degrader placed at the intermediate image plane was fundamental for the separation of the transmitted nuclei \cite{Sch87} and the identification of atomic charge states, as explained below. The nuclei traversing the spectrometer were identified by determining their magnetic rigidity, velocity and atomic number. The magnetic rigidity was derived from the positions of the nuclei along the dispersive coordinate at the intermediate and final image planes. These positions were measured with two plastic scintillators covering both image planes and providing the arrival times of the induced signals at both ends of the scintillation plate. These two scintillators provided also the time of flight of the ions between both image planes ($\approx$ 35 m). Finally, two multi-sampling ionisation chambers placed at the end of the spectrometer provided the identification of the residual nuclei in atomic number from the registered energy loss. A description of the identification method for these nuclei can be found in \cite{Cas06,Tai03}.

The unambiguous identification of neutron-rich nuclei requires the separation of the different charge states of the nuclei transmitted through the spectrometer. Those nuclei changing their charge state while traversing the matter located at the intermediate image plane (plastic scintillator and energy degrader) can be easily identified by their change in magnetic rigidity. However, the identification of non fully stripped nuclei preserving their charge state along the spectrometer is a challenging problem. Hydrogen-like nuclei have almost the same mass over ionic charge state ratio A/q as fully stripped nuclei of the same element with three more neutrons. Since the cross sections of the most neutron-rich isotope is, in average, around two orders of magnitude smaller than the contaminant, contributions of hydrogen-like nuclei of the order of few per cent represent a large contamination for the identification of the most neutron-rich isotopes. This problem is particularly important in the region of interest for this work since the fraction of non fully stripped nuclei increases considerably with the atomic number of the nuclei \cite{Sch98}.

\begin{figure*}
  \begin{center}
    \leavevmode
    \includegraphics[width=17.0cm]{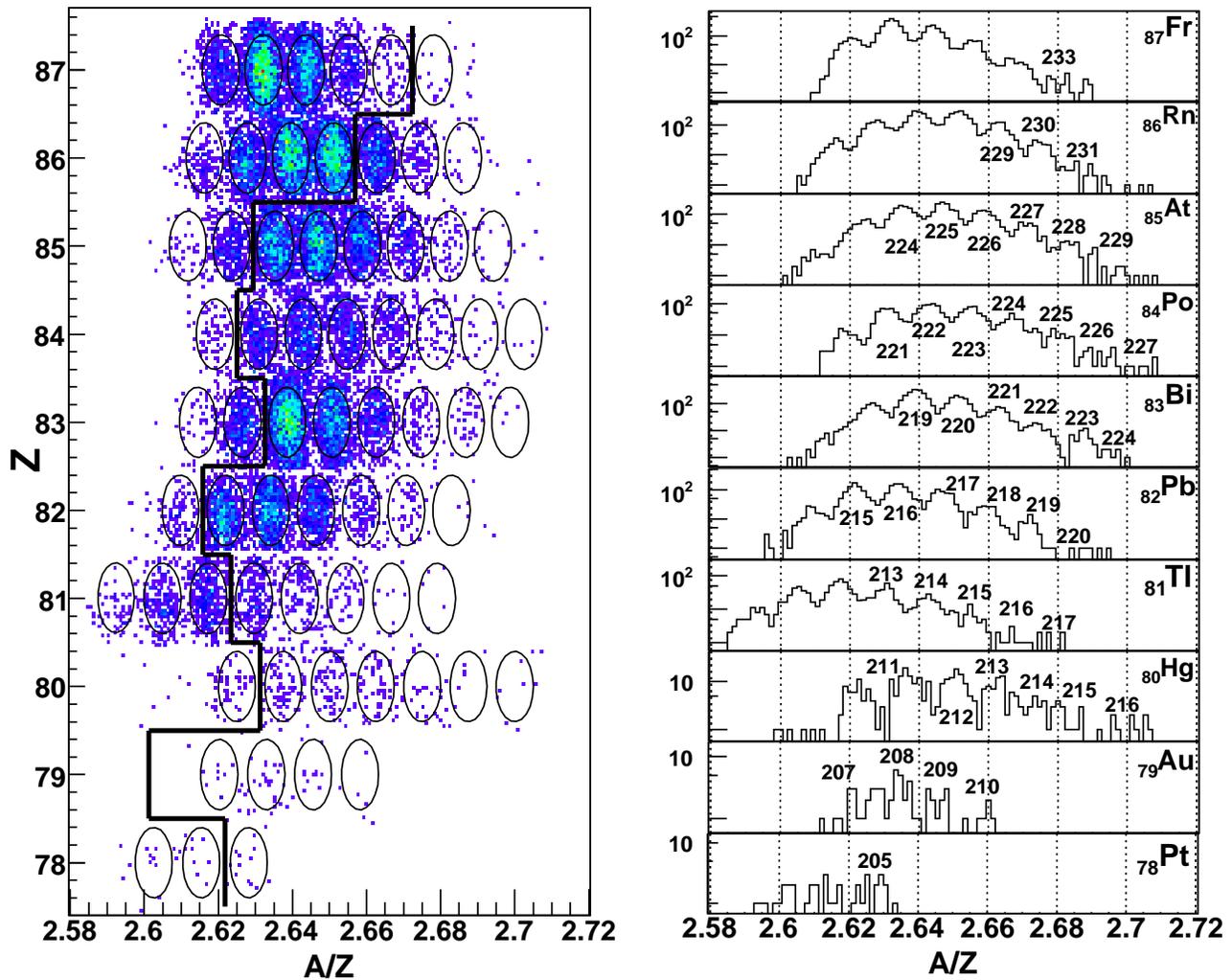}
    \caption{(Color online) Left panel: Identification plot of all nuclei produced in this work, see text for details. Right panel: A/Z distribution of the recorded events for elements between Pt and Fr. Previously unknown nuclei are indicated by their mass number.}
\label{fig_1}
  \end{center}
\end{figure*}

The key parameter in the present experiment for overcoming this problem was the relativistic energy of the nuclei under investigation, reducing drastically the fraction of non fully stripped nuclei \cite{Sch98}. For example, using a 1 A GeV $^{238}$U beam, the expected fraction of fully stripped $^{226}$Po nuclei produced in a beryllium target having a thickness equivalent to 20\% of the range of the projectile and a 120 mg/cm$^2$ niobium stripper is 89\%. At 500 A MeV that fraction would be only 58\% \cite{Aud06}. Moreover, we used two multi-sampling ionisation chambers at the final image plane with a stripper foil in between, further reducing the amount of non fully stripped nuclei. As explained in Ref. \cite{Ben99}, combining the difference in magnetic rigidity in the two sections of the spectrometer and the difference in energy loss in both ionisation chambers we could separate the charge states of the transmitted nuclei and, in particular, most of the hydrogen-like nuclei that preserve their charge state along the experimental set-up \cite{Cas06}. The final absolute contamination of the remaining hydrogen-like nuclei to the corresponding A+3 isotopes was estimated to be, at most, 40\%.

\begin{figure*}
  \begin{center}
    \leavevmode
    \includegraphics[width=18.0cm]{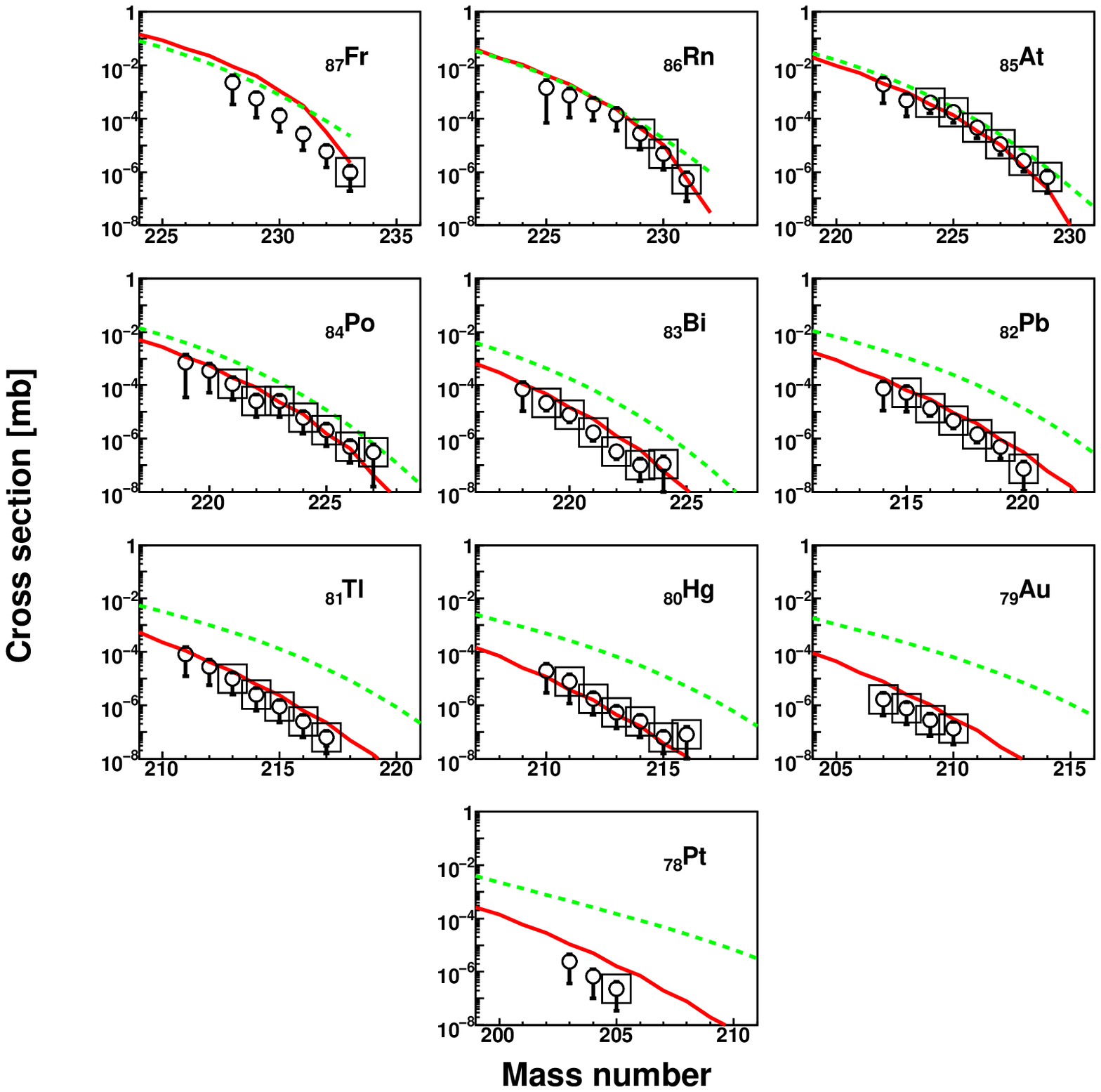}
    \caption{Isotopic distributions of the production cross sections of heavy neutron-rich nuclei determined in this work. Nuclei observed for the first time are surrounded by a square symbol. The experimental measurements are compared with the predictions obtained with the code COFRA (solid line) and EPAX (dashed line).}
\label{fig_2}
  \end{center}
\end{figure*}

In order to search for new heavy neutron-rich nuclei, we tuned the FRS magnets for centering the following nuclei, $^{227}$At, $^{229}$At, $^{216}$Pb, $^{219}$Pb and $^{210}$Au, along its central trajectory. Combining the signals recorded in these settings of the FRS and using the analysis technique previously explained we were able to identify 45 new neutron rich nuclei with atomic numbers between Z=78 and Z=87: $^{205}$Pt, $^{207-210}$Au, $^{211-216}$Hg, $^{213-217}$Tl, $^{215-220}$Pb, $^{219-224}$Bi, $^{221-227}$Po, $^{224-229}$At, $^{229-231}$Rn and $^{233}$Fr. In Fig.\ref{fig_1} we present the identification matrix obtained for all the observed isotopes of elements between platinum and francium. The criteria followed for accepting the identification of a nucleus observed for the first time is a number of events compatible with the corresponding mass and atomic number, located in the expected range of positions at both images planes of the spectrometer and with a probability of being background or contaminants below 5\%. The left panel in this figure corresponds to the scatter-plot where we represent atomic number versus the ratio mass over nuclear charge, which allows us to identify the nuclei measured in the experiment. In this figure the horizontal and vertical thick-solid lines show the present limits of the chart of nuclides. Therefore, all nuclei located to the right of these lines were previously unobserved. The production of $^{203}$Pt and $^{204}$Pt in the fragmentation of $^{208}$Pb projectiles at 1 A GeV was previously reported by some of us  \cite{Kur06}. In the right panel of the figure we represent the corresponding mass-over-nuclear charge distribution for the most neutron-rich isotopes of elements between platinum and francium we have measured. This plot clearly shows the resolving power and the unambiguous isotopic identification we obtain. We indicate the previously unknown isotopes by their mass number. 

We could also determine the production cross sections of these nuclei normalising the production yields to the integrated beam current and the number of target nuclei. The beam current was continuously measured during the experiment using a secondary-electron monitor placed before the reaction target. This monitor was carefully calibrated during the experiment using a plastic scintillator \cite{Cas06}. The measured yields were also corrected by the losses inherent to the experimental technique used in this work. The main corrections were due to the limited momentum acceptance of the spectrometer, in particular, for those nuclei with magnetic rigidities far from the central value defined by the FRS tune, the fraction of non fully stripped nuclei, and secondary reactions in all layers of matter placed along the spectrometer. Smaller corrections were due to the acquisition dead time and detectors efficiency. 

Particular attention was paid to the evaluation of the uncertainties associated to the measured cross sections. Statistical uncertainties were below 10\%, except for the two most neutron-rich nuclei measured per element with much smaller statistics. The systematic uncertainties were associated to the different corrections applied to the measured yields, and their magnitude was carefully evaluated around 20\%. 

In Fig. \ref{fig_2} we represent the isotopic distributions of the cross sections measured in this work. The error bars are shown when larger than the data points. Those points in the figure surrounded by a square correspond to the 45 new isotopes discovered in this experiment. In a few days measurement we were able to reach cross sections as low as 100 pbarn. As expected, the production cross sections decrease drastically with the neutron number. On average, an additional neutron decreases the production cross section about a factor 4.

In the same figure we also compare the measured cross sections with predictions obtained with the codes EPAX \cite{Sum00} and COFRA \cite{Ben99,Ben10}. EPAX is a well known parametrisation of measured cross sections while COFRA is an analytical version of the abrasion-ablation fragmentation model of Gaimard and Schmidt \cite{Gai91}. The COFRA code uses a full description of the abrasion stage of the collision while the ablation stage considers only the evaporation of neutrons, which is well adapted to the present case. This simplification makes possible an analytical formulation of the de-excitation stage that reduces considerably the computation time. Therefore, COFRA includes fluctuations in the number of abraded protons and neutrons according to a hyper-geometrical distribution \cite{Huf75} and in the excitation energy gained by the pre-fragments due to the random hole creation in the Fermi distribution of the nucleons inside the nucleus \cite{Ben99}.

The benchmarking of both calculations yields the following conclusions. The EPAX code describes rather well the production of residual nuclei relatively close in mass number to the projectile however, it over-predicts the production cross sections of neutron-rich residual nuclei produced in more central collisions where the projectile loses an important number of nucleons. Similar conclusions had been obtained in other works \cite{Ben08}, EPAX describes rather well the production cross sections of neutron-deficient nuclei in fragmentation reactions but it overestimates the production of neutron-rich nuclei relatively far in mass number to the initial projectile.

COFRA provides an overall good description of the cross sections of neutron-rich nuclei produced in fragmentation reactions induced by $^{238}$U projectiles. A more detailed analysis indicates a slight overestimation of the cross sections of nuclei far from the projectile. It has been indicated that this effect could be due to the depopulation of these neutron-rich nuclei by fission reactions not included in the COFRA code \cite{Alv09}. However, the fact that this slight overestimation of the cross sections of neutron-rich residual nuclei in more central collisions has also been observed in the fragmentation of non fissile projectiles such as $^{136}$Xe \cite{Ben08} rules out this conclusion. 

In this work we have investigated the production of heavy neutron-rich nuclei using fragmentation reactions of relativistic $^{238}$U projectiles. The use of an intense beam and a high-resolving-power magnetic spectrometer made it possible to produce in few days a large number of neutron-rich isotopes of elements between platinum and francium with cross sections as low as 100 pbarn. In particular we identified for the first time 45 new neutron-rich isotopes of these elements, $^{205}$Pt, $^{207-210}$Au, $^{211-216}$Hg, $^{213-217}$Tl, $^{215-220}$Pb, $^{219-224}$Bi, $^{221-227}$Po, $^{224-229}$At, $^{229-231}$Rn and $^{233}$Fr. It was also shown that a key parameter for the unambiguous identification of these new nuclei was the high energy of the projectiles, reducing drastically the contamination of atomic charge states. The production cross sections of these nuclei were also determined with good accuracy. These cross sections were used to benchmark the two most widely used codes for estimating the production cross sections of neutron-rich residual nuclei in fragmentation reactions. While EPAX clearly overestimate the production of residual nuclei far in mass number to the initial projectile, the COFRA code provides an overall good description of the cross sections. The description of the data with this code confirms the large fluctuations in the number of abraded protons and neutrons and in the excitation energy gained by the prefragments. This large fluctuations make possible to populate cold-fragmentation reactions channels leading to the production of the most neutron-rich nuclei. These results pave the way for a considerable extension of the north-east limit of the chart of nuclides expected with the new generation of radioactive beam facilities.

\begin{acknowledgments}

This work was partially supported  by the European Community under the FP6 ``Integrated Infrastructure Initiative EURONS'' contract n. RII3-CT-2004-506065, by the Spanish Ministry of Science and Innovation under grants FPA2007-62652 and the programme ``Ingenio 2010, Consolider CPAN'', and by ``Xunta de Galicia'' under grant PGIDT00PXI20606PM. The EC is not liable for any use that may be made of the information contained herein.

\end{acknowledgments}


\end{document}